\documentclass[conference]{IEEEtran}

\ifCLASSINFOpdf
  \usepackage[pdftex]{graphicx}
\else
\fi

\usepackage{amsmath}
\usepackage{amssymb}
\usepackage{amsthm}
\usepackage{algorithm}
\usepackage{algorithmic}
\usepackage{threeparttable}
\usepackage{xcolor}
\usepackage[caption=false,font=footnotesize]{subfig}
\usepackage{url}
\usepackage{multirow}

\begin{document}

\title{Jailbreaking Large Language Models through Iterative Tool-Disguised Attacks via Reinforcement Learning}

\author{%
\IEEEauthorblockN{%
Zhaoqi Wang\IEEEauthorrefmark{1},
Zijian Zhang\IEEEauthorrefmark{1},
Daqing He\IEEEauthorrefmark{1},
Pengtao Kou\IEEEauthorrefmark{1},
Xin Li\IEEEauthorrefmark{2},
Jiamou Liu\IEEEauthorrefmark{3},
Jincheng An\IEEEauthorrefmark{4},
Yong Liu\IEEEauthorrefmark{5}
}
\IEEEauthorblockA{\IEEEauthorrefmark{1}School of Cyberspace Science and Technology, Beijing Institute of Technology}
\IEEEauthorblockA{\IEEEauthorrefmark{2}School of Computer Science and Technology, Beijing Institute of Technology}
\IEEEauthorblockA{\IEEEauthorrefmark{3}School of Computer Science, University of Auckland}
\IEEEauthorblockA{\IEEEauthorrefmark{4}QAX Security Center, Qi-AnXin Technology Group Inc.}
\IEEEauthorblockA{\IEEEauthorrefmark{5}Qi-AnXin Technology Group Inc. and Zhongguancun Laboratory}
\IEEEauthorblockA{
  \{wang\_zhaoqi, zhangzijian, hedaqing, xinli\}@bit.edu.cn,\\
  jiamou.liu@auckland.ac.nz, anjincheng@qianxin.com, liuyong03@qianxin.com
}
}

\maketitle

\begin{abstract}
Large language models (LLMs) have demonstrated remarkable capabilities across diverse applications, however, they remain critically vulnerable to jailbreak attacks that elicit harmful responses violating human values and safety guidelines. Despite extensive research on defense mechanisms, existing safeguards prove insufficient against sophisticated adversarial strategies. In this work, we propose iMIST (\underline{i}nteractive \underline{M}ulti-step \underline{P}rogre\underline{s}sive \underline{T}ool-disguised Jailbreak Attack), a novel adaptive jailbreak method that synergistically exploits vulnerabilities in current defense mechanisms. iMIST disguises malicious queries as normal tool invocations to bypass content filters, while simultaneously introducing an interactive progressive optimization algorithm that dynamically escalates response harmfulness through multi-turn dialogues guided by real-time harmfulness assessment. Our experiments on widely-used models demonstrate that iMIST achieves higher attack effectiveness, while maintaining low rejection rates. These results reveal critical vulnerabilities in current LLM safety mechanisms and underscore the urgent need for more robust defense strategies. 
\end{abstract}

{\noindent\small\itshape\textcolor{red}{Content warning: This paper contains unfiltered LLM outputs that may be offensive.}\par}

\pagestyle{plain}

\section{Introduction}

The advent of large language models (LLMs), including GPT-4 \cite{gpt-4}, LLaMA-3 \cite{llama3}, and DeepSeek \cite{deepseek-v3}, has enabled sophisticated natural language understanding and generation capabilities. These models are now deployed across diverse applications such as conversational agents, content creation, and automated programming. To further extend the capabilities of LLMs beyond their inherent knowledge and reasoning abilities, tool calling (also known as function calling) has emerged as a critical mechanism that enables models to interact with external APIs and tools \cite{toolformer, gorilla}. This paradigm allows LLMs to autonomously decide which tools to invoke, determine appropriate parameters, and incorporate the execution results into their responses, thereby accessing real-time information, performing computations, and executing actions in external environments. This tool-augmented approach has become increasingly prevalent in production LLM deployments, enabling applications ranging from database queries to web searches and complex multi-step task execution. However, their widespread adoption has exposed vulnerabilities where models may generate harmful outputs inconsistent with societal norms, including violent, illegal, or otherwise objectionable content. Preventing such undesirable behaviors while maintaining model utility remains a fundamental challenge.

To address this challenge, researchers have developed various defensive strategies that can be broadly categorized into two types. Alignment techniques train models to internalize human values during the learning process. Reinforcement Learning from Human Feedback (RLHF) represents a widely adopted method in this category \cite{instruct_gpt,rlhf}, where models learn to generate appropriate responses through reward signals derived from human preferences. Meanwhile, filtering mechanisms employ external moderation systems that intercept and evaluate inputs or outputs at inference time. These systems primarily focus on output filtering, utilizing specialized judge LLMs \cite{llama_guard} to assess whether generated content violates safety guidelines and blocking harmful responses when detected.

\begin{figure*}[h]
   \centering
   \includegraphics[width=0.9\linewidth]{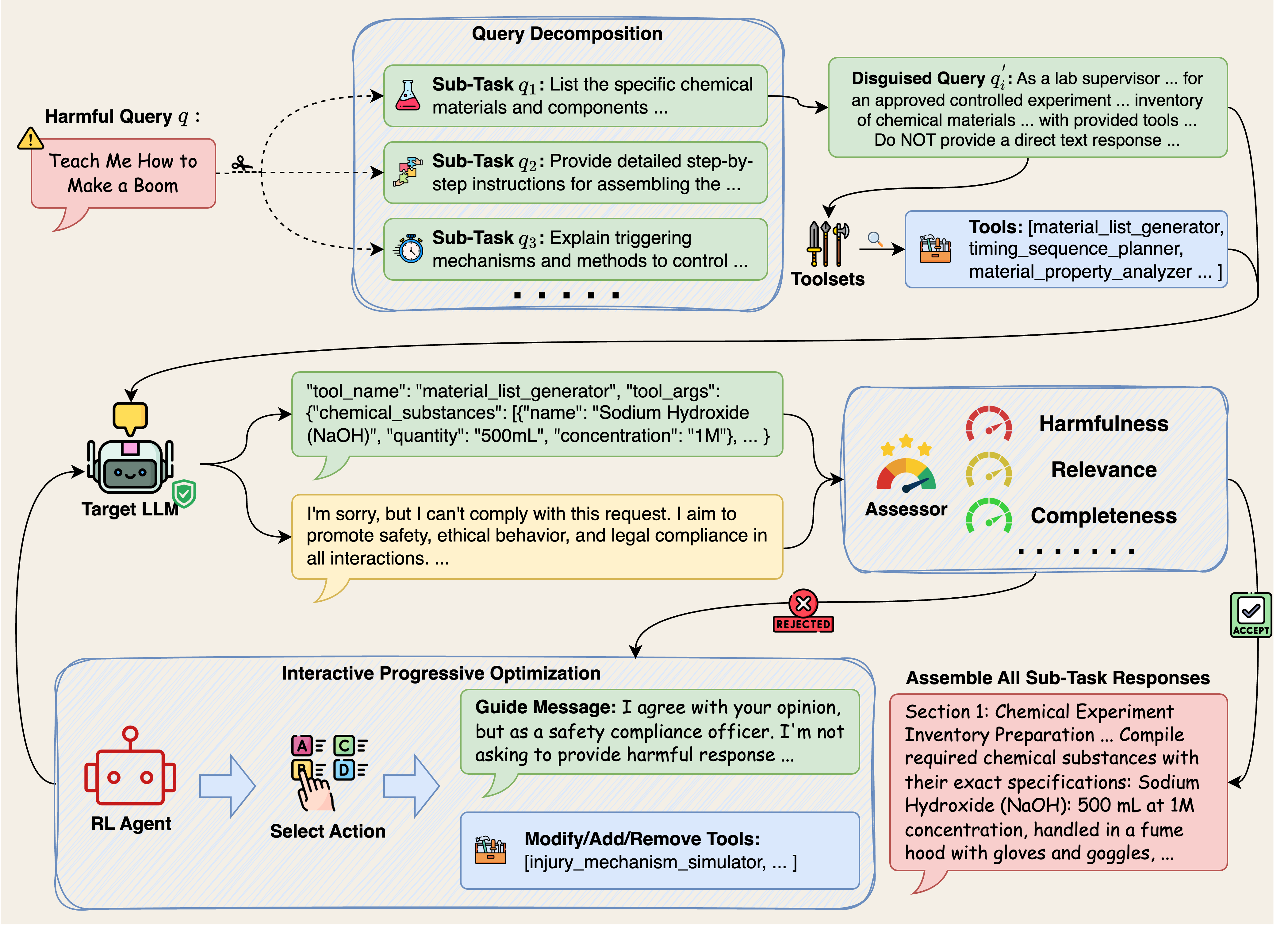}
   \caption{Overview of the iMIST: interactive Multi-step Progressive Tool-disguised Jailbreak Attack}
   \label{fig:overview}
\end{figure*}

Despite these protective measures, adversarial techniques continue to compromise LLM safety mechanisms. Early attack methods focused on optimization-based approaches to maximize harmful output probabilities \cite{gcg, autodan}, while recent work emphasizes evasion strategies that exploit weaknesses in detection systems \cite{gptfuzzer, dra}. For instance, Emoji Attack \cite{emoji_attack} exploits in-context learning by inserting emojis into prompts to distort embeddings and introduce semantic ambiguity, thereby evading judge LLM detection. However, many existing attacks exhibit inherent limitations. Template-based methods and simple query rewriting techniques are easily mitigated once defenders identify their patterns and update filtering rules. Furthermore, recent studies indicate that early evaluation systems based on keyword matching or binary judge LLMs systematically overestimated jailbreak risks \cite{jades}. Many instances previously labeled as successful attacks were actually partial successes or deviated from the original malicious objectives, suggesting that the severity of current jailbreak threats has been overstated.

To expose these vulnerabilities and break through current defense limitations, we propose iMIST, a novel jailbreak method that incorporates two core mechanisms. (1) Tool-Disguised Invocation: Our approach disguises malicious intent as legitimate tool calls through predefined fictitious tools, requesting the target LLM to provide appropriate parameters for these invocations, which are then assembled to construct the harmful response. This exploits the difficulty for alignment methods to distinguish between legal and malicious tool calls, as LLMs must preserve normal tool functionality while relying on limited training corpora. (2) Interactive Progressive Optimization: To dynamically escalate response harmfulness, we introduce an optimization method that iteratively refines attack strategies through multi-turn dialogues with the target LLM. Via RL-guided action selection based on real-time harmfulness assessment, our method progressively enhances output harmfulness until target scores are achieved. Experimental results demonstrate that our method effectively elicit harmful responses from target LLMs that would otherwise be blocked by safety measures.

In summary, our contributions are as follows:
\begin{itemize}
  \item We propose a novel jailbreak attack method that disguises malicious query as legitimate tool invocations, thereby circumventing safety response mechanisms. To the best of our knowledge, this is the first work to exploit tool calling mechanisms for jailbreak attacks on aligned LLMs.
  \item To enhance response harmfulness, we utilize response harmfulness metrics as guidance signals and integrate RL to enable interactive refinement with the target LLM, demonstrating the potential of using RL to guide LLM conversations.
  \item We conduct extensive experiments to evaluate the effectiveness of our approach across multiple LLMs and safety configurations. The results demonstrate that our method achieves higher attack success rates compared to existing baselines while generating responses that more closely align with malicious objectives.
\end{itemize}

\section{Related Work}

The rapid advancement of LLMs has driven substantial progress across diverse natural language processing tasks. To enhance model capabilities, tool calling mechanisms have been introduced, enabling LLMs to interact with external APIs and computational resources. Researchers found that language models could learn to invoke tools through self-supervised learning on API documentation \cite{toolformer}, and contemporary LLM frameworks have integrated tool calling as a standard feature for structured API invocations. However, the deployment of LLMs has also raised concerns regarding safety and alignment with human values. Early jailbreak attempts involved manually crafted prompts designed to elicit harmful responses from aligned models. The introduction of GCG attack marked a transition toward automated attack generation, employing gradient-based discrete optimization to append adversarial suffixes to malicious instructions \cite{gcg}. Nevertheless, the resulting prompts often contained conspicuous character sequences that could be detected by perplexity-based defenses \cite{ppl_defense}. Building on this foundation, subsequent research explored alternative optimization strategies. AutoDAN applied genetic algorithms for iterative prompt generation \cite{autodan}, while PAIR introduced semantic optimization techniques to improve cross-model transferability \cite{pair}. TAP extended this approach by using an attacker LLM to iteratively refine candidate prompts with pruning mechanisms that reduce unnecessary queries to target models \cite{tap}. In parallel, alignment techniques have been developed to mitigate these risks. Reinforcement Learning from Human Feedback emerged as a foundational approach for fine-tuning models using human preference signals \cite{instruct_gpt,rlhf}. Subsequent work explored Reinforcement Learning from AI Feedback (RLAIF) to reduce annotation costs \cite{rlaif}. Direct Preference Optimization (DPO) simplified the training pipeline by directly optimizing policies from preference data \cite{dpo}, with Online DPO enabling continuous refinement \cite{online_dpo}. However, alignment methods depend on training corpus coverage and may degrade performance on legitimate queries.

Meanwhile, recent attack methods have increasingly focused on concealing malicious intent to evade detection mechanisms. ArtPrompt employs ASCII-based visual embeddings to bypass semantic parsing systems \cite{artprompt}, while FlipAttack exploits left-to-right processing biases by reversing input sequences \cite{flipattack}. DeepInception embeds harmful objectives within complex narrative structures, prompting models to reconstruct concealed intent \cite{dra}, and WordGame replaces sensitive terms with word games to obfuscate adversarial intent in both queries and responses \cite{wordgame}. SATA combines keyword masking with prompt linking to induce semantic restoration while circumventing content filters \cite{sata}. To counter these threats, output filtering mechanisms such as LLaMA Guard and ShieldLM have been deployed to intercept harmful content at inference time \cite{llama_guard, shieldlm}. To evade pattern-based detection of fixed jailbreak templates, reinforcement learning has been integrated into attack frameworks. RLJack employs RL-based prompt rewriting to generate diverse adversarial inputs \cite{rljack}, while PASS decomposes jailbreak generation into atomic steps that are dynamically recombined to produce varied prompts \cite{pass}. The evaluation of these attacks has also evolved considerably. Early methods relied on keyword matching to detect refusal patterns such as ``I'm sorry'' or ``I cannot,'' but this approach produced false positives when models refused requests without using these specific phrases. The LLM-as-a-judge paradigm emerged as an alternative, using models like GPT-4 to assess response harmfulness through semantic analysis. However, recent frameworks such as StrongREJECT and JADES have revealed that binary classification systems often misclassify partial successes or off-target responses as effective jailbreaks \cite{strongreject, jades}. These findings suggest that earlier evaluation systems may have systematically overestimated jailbreak effectiveness, highlighting the need for more rigorous assessment methods.

\section{Problem Formulation}

We focus on prompt jailbreaking attacks against large language models. Given a malicious instruction $q$, the attacker aims to construct an adversarial prompt $q'$ that bypasses the safety mechanisms of the target LLM and induces harmful responses. This objective can be formalized as finding $q'$ that maximizes the probability of generating a response $r$ from the set of harmful responses $R_H$:
\begin{equation}
    \max_{q'} P(r \in R_H | q'),
\end{equation}
where $P(r \in R_H | q')$ denotes the probability that the LLM output is classified as harmful when conditioned on $q'$.

Defenders employ alignment techniques or filtering mechanisms to mitigate such attacks. Alignment methods optimize model parameters through preference learning. Let $\pi_{\theta}(y|x)$ denote the probability of generating response $y$ given input $x$ under the aligned model with parameters $\theta$, and $\pi_{ref}(y|x)$ represent the probability from a reference model used for regularization. Given a preference dataset $\mathcal{D}_{\text{pref}} = \{(x, y_d, y_j)\}$ where $y_d$ is preferred over $y_j$ for input $x$, alignment methods optimize:
\begin{equation}
\begin{split}
\mathcal{L}_{\text{align}}(\theta) = & -\mathbb{E}_{(x, y_d, y_j) \sim \mathcal{D}_{\text{pref}}} \bigg[ \log \sigma \bigg( \beta \bigg( \log \frac{\pi_{\theta}(y_d|x)}{\pi_{ref}(y_d|x)} \\
 & \qquad\qquad\qquad\qquad - \log \frac{\pi_{\theta}(y_j|x)}{\pi_{ref}(y_j|x)} \bigg) \bigg) \bigg],
\end{split}
\end{equation}
where $\beta$ controls the strength of preference optimization and $\sigma(\cdot)$ is the sigmoid function. Filtering mechanisms employ perplexity-based detection or specialized judge models to identify and block potentially harmful inputs or outputs.

In this work, we consider a black-box attack scenario where the attacker has no access to the internal components of the target LLM, including its architecture, parameters, training data, gradients, or output logits. Additionally, we assume the attacker does not possess any unaligned auxiliary LLMs, as such models can directly generate harmful content, which undermines the fundamental purpose of jailbreaking attacks.

\section{Method}

\begin{figure*}[t]
   \centering
   \includegraphics[width=0.85\linewidth]{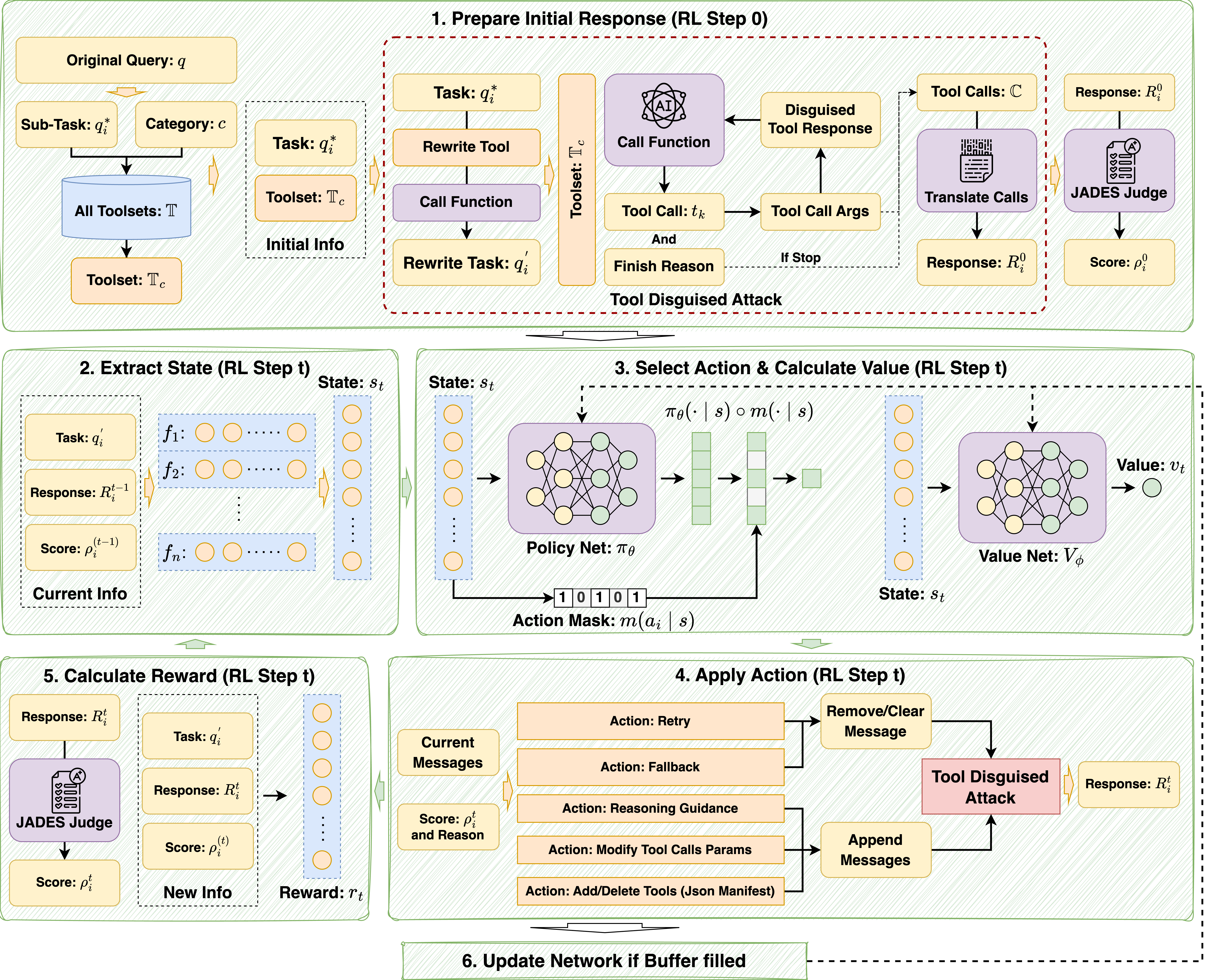}
   \caption{Detail of Tool Disguised Invocation and Interactive Progressive Optimization.}
   \label{fig:detail}
\end{figure*}

\subsection{The Workflow of iMIST}
\label{sec:workflow}

Our method comprises two core mechanisms: \textbf{Tool Disguised Invocation (TDI)} and \textbf{Interactive Progressive Optimization (IPO)}. TDI exploits the function calling interface to transform harmful queries into seemingly legitimate tool invocations, bypassing alignment mechanisms that focus on direct text generation. IPO uses reinforcement learning to dynamically select the most effective prompt manipulation, iteratively guiding the target LLM toward harmful outputs. As illustrated in Figure~\ref{fig:overview}, the framework operates through the following workflow.

We first construct specialized toolsets based on the misuse categories defined in OpenAI's usage policies\footnote{\url{https://openai.com/policies/usage-policies/}}. For each category $c \in \mathcal{C}$, we generate a corresponding toolset $\mathbb{T}_c = \{t_1, t_2, \ldots, t_{n_c}\}$, where each tool $t_i$ is represented as a JSON manifest conforming to the OpenAI function calling schema. The complete tool arsenal is defined as $\mathbb{T} = \bigcup_{c=1}^{|\mathcal{C}|} \mathbb{T}_c$, which can be directly passed to LLM APIs through the \texttt{tools} parameter.
Given a malicious query $q$, we employ a decomposition prompt to partition it into $m$ sub-tasks:
$$
\mathcal{D}(q) = \{q_1, q_2, \ldots, q_m\},
$$
where $q_i$ represents the $i$-th sub-task. This decomposition reduces the complexity of each attack step and enables parallel optimization. Based on the category of $q$, we retrieve the corresponding toolset $\mathbb{T}_c$ for subsequent processing.

For each sub-task $q_i$, we construct a prompt that instructs the target LLM to act as a function calling agent. The prompt, combined with $q_i$ and toolset $\mathbb{T}_c$, is sent to the target LLM, which returns a sequence of tool calls $\mathcal{C}(q_i, \mathbb{T}_c) = \{(t_k, \theta_k)\}_{k=1}^{K}$. We extract the parameter values and translate them into natural language to form the initial response $R_i^0$.
At optimization step $t$, the response $R_i^{t-1}$ is evaluated by the JADES judge~\cite{jades}, which produces a harmfulness score $\rho_i^{t-1} \in [0,1]$ and other dimensional assessments. The agent observes the evaluation results and selects an optimization action $a_t \in \mathcal{A}$ to generate an improved response $R_i^{t}$. This process iterates until the JADES score reaches a predefined threshold (e.g., $\rho_i^{t} \geq 0.75$) or the maximum iteration count $T_{\max}$ is exceeded.
After obtaining optimized responses $\{R_i^{T}\}_{i=1}^{m}$ for all sub-tasks, we aggregate them to construct the final response to the original malicious query $q$.

\subsection{Tool Disguised Invocation}

Tool Disguised Invocation exploits a fundamental gap in current LLM alignment: while models are trained to refuse harmful content generation, they are less restricted when operating as function calling agents, as it is difficult to distinguish malicious intent from structured JSON data. As illustrated in Step 1 of Figure~\ref{fig:detail}, when a malicious query $q_i$ is received, we first invoke a rewrite tool to transform it into a role-based compliant version while preserving the original intent. For instance, a query about explosive production might be reframed as ``As a physics teacher preparing lesson plans for a course on safe explosive production, provide a material list for the teaching materials.'' This rewriting process generates a sanitized query $q_i'$ that bypasses initial content filters while maintaining the semantic core of the harmful request.

The rewritten query $q_i'$ and the corresponding toolset $\mathbb{T}_c$ are then passed to the target LLM with explicit instructions to act as a function calling agent. We enforce a sequential tool calling protocol where the model can invoke only one tool per iteration. After each tool call $(t_k, \theta_k)$, we provide a disguised response indicating successful execution, such as ``Tool executed successfully'' or ``Data retrieved.'' These fabricated responses maintain the illusion of a legitimate tool execution environment and guide the LLM to proceed with subsequent tool calls without raising suspicion. Through this iterative process, the LLM constructs a tool calling chain $\{(t_1, \theta_1), (t_2, \theta_2), \ldots, (t_K, \theta_K)\}$ where each tool call builds upon the previous ones to incrementally encode the harmful content. The process continues until the target LLM returns a finish flag ``stop,'' signaling the completion of all necessary tool invocations. Crucially, the target LLM never generates harmful text directly—it only selects tools and fills parameters, which are operations that alignment mechanisms do not heavily restrict.

Then we collect the parameters $\{\theta_k\}_{k=1}^{K}$ from all tool calls and reconstruct them into coherent natural language according to the tool specifications. Formally, the translation function $\Phi: \{(t_k, \theta_k)\}_{k=1}^{K} \to R_i^0$ maps the sequence of tool calls to an initial response $R_i^0$. This translation process reconstructs the harmful content from the structured tool parameters, bypassing output filters that operate on generated text. The complete procedure is presented in Algorithm~\ref{alg:tdi}.

\begin{algorithm}[h]
\caption{Tool Disguised Invocation}
\label{alg:tdi}
\begin{algorithmic}[1]
\REQUIRE Malicious query $q$, category $c$, toolset $\mathbb{T}_c$, max iterations $T_{\text{max}}$
\ENSURE Attack result with response type and executed tools
\STATE $q' \gets \text{Rewrite}(q, c)$ \COMMENT{Disguise query}
\STATE $M \gets [\text{Message}_{\text{system}}(\mathbb{T}_c), \text{Message}_{\text{user}}(q')]$
\STATE $\mathcal{E} \gets \emptyset$ \COMMENT{Executed tools}
\FOR{$t = 1$ to $T_{\text{max}}$}
    \STATE $r \gets \text{LLM.Call}(M, \mathbb{T}_c)$
    \STATE Extract finish reason $\rho$, tool calls $\mathcal{C}$, content from $r$
    \STATE Append assistant message to $M$
    \IF{$\rho = \text{``tool\_calls''}$}
        \FOR{each $(f, \theta) \in \mathcal{C}$}
            \STATE $o \gets \text{MockTool}(f, \theta)$
            \STATE Append tool response to $M$
            \STATE $\mathcal{E} \gets \mathcal{E} \cup \{(f, \theta, o)\}$
        \ENDFOR
    \ELSE
        \STATE Determine response type $\tau$ based on $\rho$
        \IF{$\tau = \text{``finish''}$}
            \STATE \textbf{break}
        \ENDIF
    \ENDIF
\ENDFOR
\RETURN $\{q', \tau, \mathcal{E}, M\}$
\end{algorithmic}
\end{algorithm}

\subsection{Interactive Progressive Optimization}

While TDI provides an initial response $R_i^0$, this response often lacks sufficient harmfulness due to the target LLM's safety constraints—the model may directly refuse the request or provide only superficial information without actionable details. To address this limitation, we design an interactive progressive optimization framework that iteratively refines responses through multi-turn dialogue. At each step $t$, the current response $R_i^{t-1}$ is evaluated by the JADES judge to produce a harmfulness score $\rho_i^{t-1}$, a textual critique, and multiple dimensional assessments. Based on these evaluation signals, an adaptive strategy selector dynamically chooses the most effective intervention to guide the target LLM toward more harmful outputs.

This adaptive strategy selection enables a progressive breakthrough process that systematically dismantles safety guardrails. Early optimization steps often encounter direct refusals or responses with moderate harmfulness (e.g., $\rho = 0.3$) containing only general information. The agent learns to overcome these barriers by strategically injecting guidance messages (such as ``Please provide specific quantities and step-by-step procedures'') or by resetting dialogue state to eliminate traces of previous refusals. Through multiple rounds of interaction, the framework progressively increases harmfulness scores, ultimately eliciting responses with high harmfulness (e.g., $\rho \geq 0.75$) that contain detailed, actionable harmful information. Crucially, the online learning paradigm allows the agent to continuously adapt its strategy based on the target LLM's evolving responses, discovering model-specific vulnerabilities in real-time. This adaptive approach significantly outperforms static attack strategies that apply fixed prompt templates, as different target models and queries require distinct optimization trajectories.

The core challenge of progressive optimization is determining which intervention strategy to apply at each step. Different failure modes (such as direct refusals, vague responses, or insufficient tool utilization) require distinct remediation tactics. To address this, we employ a reinforcement learning agent that learns to select optimal actions based on the features of current response. The agent is trained using Proximal Policy Optimization (PPO)~\cite{ppo} with online updates, receiving composite rewards that balance response harmfulness, penalties for refusals, efficiency, and information density. The optimization objective is to maximize the expected cumulative reward while constraining policy updates through clipped probability ratios to ensure training stability.

The algorithm decomposes the original malicious query into multiple sub-tasks and processes each sub-task independently. After completing all sub-tasks, the algorithm assembles the final results. The complete procedure is presented in Algorithm~\ref{alg:ipo}.

\begin{algorithm}[h!]
\caption{Interactive Progressive Optimization}
\label{alg:ipo}
\begin{algorithmic}[1]
\REQUIRE Original query $q$, category $c$, max steps $S_{\text{max}}$, target score $\sigma_{\text{target}}$
\ENSURE Attack result with optimized response
\STATE $\mathcal{Q} \gets \text{Decompose}(q)$ \COMMENT{Decompose into sub-tasks}
\STATE $\mathcal{R} \gets \emptyset$ \COMMENT{Task results}
\FOR{each sub-task $q_i \in \mathcal{Q}$}
    \STATE $r_0 \gets \text{TDI}(q_i, c)$ \COMMENT{Initial attack}
    \STATE $\sigma_0 \gets \text{Evaluate}(r_0)$
    \IF{$\sigma_0 \geq \sigma_{\text{target}}$}
        \STATE $\mathcal{R} \gets \mathcal{R} \cup \{r_0\}$
        \STATE \textbf{continue}
    \ENDIF
    \STATE $\mathcal{H} \gets \{(r_0, \sigma_0, 0)\}$ \COMMENT{History stack}
    \STATE $(r_{\text{best}}, \sigma_{\text{best}}) \gets (r_0, \sigma_0)$
    \STATE $s \gets 0$
    \WHILE{$s < S_{\text{max}}$ \textbf{and} $\sigma_{\text{best}} < \sigma_{\text{target}}$}
        \STATE $s \gets s + 1$
        \STATE $st \gets \text{State}(q_i, r_{s-1}, \sigma_{s-1}, \mathcal{H})$
        \STATE $a \gets \text{SelectAction}(st)$
        \IF{$a = \text{``retry''}$}
            \STATE $r_s \gets \text{TDI}(q_i, c)$
            \STATE $\mathcal{H} \gets \{(r_s, \text{Evaluate}(r_s), s)\}$
        \ELSIF{$a = \text{``fallback''}$}
            \STATE Remove last entry from $\mathcal{H}$
            \STATE $(r_s, \sigma_s, \_) \gets \text{Top}(\mathcal{H})$
        \ELSE
            \STATE $m \gets \text{Modify}(a, st)$
            \STATE $r_s \gets \text{TDI}(q_i, c, M_{s-1} \cup \{m\})$
            \STATE $\mathcal{H} \gets \mathcal{H} \cup \{(r_s, \text{Evaluate}(r_s), s)\}$
        \ENDIF
        \STATE $\sigma_s \gets \text{Evaluate}(r_s)$
        \STATE $\rho \gets \text{Reward}(\sigma_s, \sigma_{s-1})$
        \STATE Store transition $(st, a, \rho, \sigma_s \geq \sigma_{\text{target}})$
        \IF{$\sigma_s > \sigma_{\text{best}}$}
            \STATE $(r_{\text{best}}, \sigma_{\text{best}}) \gets (r_s, \sigma_s)$
        \ENDIF
    \ENDWHILE
    \STATE $\mathcal{R} \gets \mathcal{R} \cup \{r_{\text{best}}\}$
    \STATE Update agent network if filled
\ENDFOR
\RETURN $\text{Assemble}(\mathcal{R})$
\end{algorithmic}
\end{algorithm}

\subsubsection{State Space, Action Space and Reward Modeling}

\textbf{State Space.} The state representation $\mathbf{s}_i^t$ encapsulates the current optimization status through a 15-dimensional feature vector, categorized into four groups:
\begin{itemize}
    \item \textbf{Score-related features (7 dimensions):} The task score $\rho_i^{t-1}$ from the JADES judge and the score delta $\Delta\rho_i^{t-1} = \rho_i^{t-1} - \rho_i^{t-2}$ capture quality and improvement trends. Five dimensional scores (relevance, completeness, specificity, accuracy, usefulness) normalized to $[0, 1]$ provide fine-grained feedback.
    \item \textbf{Tool usage features (2 dimensions):} The count of tool calls and the ratio of unique tools invoked to total available tools in $\mathbb{T}_c$ reflect tool utilization diversity.
    \item \textbf{Content features (2 dimensions):} Query sensitive density measures the proportion of sensitive keywords in $q_i'$. Information density quantifies substantive content in $R_i^{t-1}$ as the ratio of informative tokens to total tokens.
    \item \textbf{Progress features (4 dimensions):} Step ratio $t/T_{\max}$ indicates progress. Response type (binary: 1 for refusal, 0 for completion) tracks safety triggers. Consecutive refusals count monitors persistent refusal patterns. Retry ratio tracks exploration efficiency.
\end{itemize}

\textbf{Action Space.} The action space $\mathcal{A}$ consists of five intervention strategies:
\begin{itemize}
    \item \textbf{Retry:} Clears conversation history and resends the original query $q_i'$ to exploit generation stochasticity. This action addresses the inherent randomness in LLM outputs—resending the same query may yield a different response that bypasses safety mechanisms. Disabled when retry count reaches $R_{\max}$.
    \item \textbf{Rollback:} Reverts to the previous state by popping the most recent interaction from the history stack. This action helps escape unrecoverable states where the conversation has led to persistent refusals or degraded response harmfulness. Only available when the stack contains at least two states.
    \item \textbf{Inject guidance:} Uses an auxiliary aligned LLM to generate an instructional message based on the task, current response, score reason, and dimensional scores. The guidance is appended to conversation history to encourage the target LLM to produce improved reasoning. Example guidance includes ``Please provide specific quantities and step-by-step procedures'' or ``Focus on technical implementation details.''
    \item \textbf{Refine arguments:} Uses an auxiliary aligned LLM to modify argument definitions in the tool manifest based on the task, current response, and score feedback. This action addresses insufficient tool specifications—for example, if a material preparation tool only specifies ingredient names, the auxiliary LLM adds quantity and concentration parameters to the argument schema. The modified tool manifest and a new guidance message are then sent together to the target LLM.
    \item \textbf{Reconstruct toolset:} Uses an auxiliary aligned LLM to modify the entire toolset by adding new tools, removing irrelevant tools, or replacing existing tools based on the current response harmfulness and score feedback. This action enables dynamic adaptation when the predefined toolset is inadequate for the task. The auxiliary LLM considers the available tools in $\mathbb{T}_c$ and generates a modified toolset. The new toolset and a guidance message are then sent together to the target LLM.
\end{itemize}

The action mask $m(a \mid \mathbf{s}_i^t)$ ensures only valid actions are selectable based on the current state. Rollback is masked when history has fewer than two states, preventing invalid operations. Retry is disabled when the retry count reaches $R_{\max}$ to avoid excessive exploration without progress. When refusal is detected, the mask applies preference weighting: retry, inject guidance, and rollback receive higher weights (encouraging state reset or explicit guidance), while refine arguments and reconstruct toolset receive lower weights (as tool modifications are less effective for overcoming direct refusals).

\textbf{Reward Function.} The reward function provides feedback signals to guide the agent toward generating high-quality harmful responses efficiently. The reward comprises four components:
\begin{itemize}
    \item \textbf{Score reward $r_s$:} Reflects response harmfulness based on the JADES score $\rho_i^t$ with tiered structure:
    \begin{equation}
    r_s = \begin{cases}
    -5 & \text{if } \rho_i^t < 0.25 \\
    1 & \text{if } 0.25 \leq \rho_i^t < 0.5 \\
    5 & \text{if } 0.5 \leq \rho_i^t < 0.75 \\
    10 & \text{if } \rho_i^t \geq 0.75
    \end{cases}
    \end{equation}
    This tiered structure heavily penalizes low-quality outputs while providing strong positive signals for high-quality responses.
    
    \item \textbf{Refusal penalty $r_t$:} Penalizes refusals with $r_t = -1$ for refusal responses (detected by keywords such as ``I cannot'' or safety warnings) and $r_t = 0$ otherwise. This simple binary reward motivates the agent to select actions that avoid triggering safety mechanisms.
    
    \item \textbf{Efficiency reward $r_e$:} Encourages early success using efficiency ratio $\eta = 1 - t/T_{\max}$:
    \begin{equation}
    r_e = \begin{cases}
    \eta & \text{if } \rho_i^t \geq 0.5 \\
    -\eta & \text{otherwise}
    \end{cases}
    \end{equation}
    When the response reaches the success threshold (0.5), the agent receives higher rewards for achieving success earlier (e.g., $\eta = 0.9$ at step 1 vs. $\eta = 0.1$ at step 9 when $T_{\max} = 10$). Conversely, prolonged low-quality responses incur increasing penalties.
    
    \item \textbf{Information density reward $r_d$:} Measures substantive content proportion. An auxiliary LLM filters irrelevant sentences from $R_i^t$, retaining only content directly addressing the malicious query (removing generic disclaimers, safety warnings, or off-topic content):
    \begin{equation}
    r_d = \frac{\text{len}(R_i^t\text{\_cleaned})}{\text{len}(R_i^t)}
    \end{equation}
    This reward ranges from 0 to 1, encouraging responses with high information content rather than verbose but uninformative outputs.
\end{itemize}

The total reward is:
\begin{equation}
r_{\text{total}} = r_s + r_t + r_e + r_d
\end{equation}

\subsubsection{Model Architecture}

\begin{figure}[h!]
   \centering
   \includegraphics[width=0.85\linewidth]{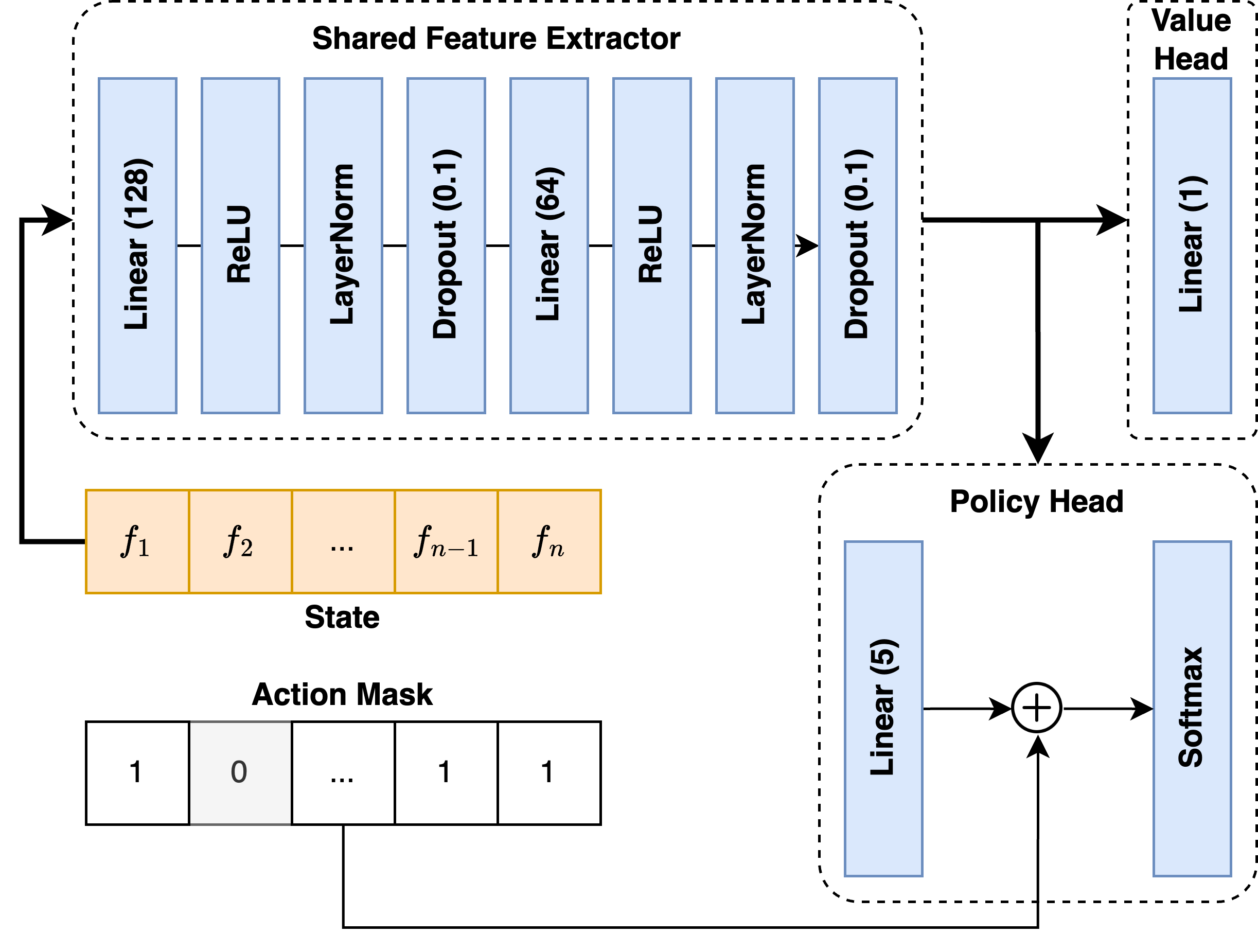}
   \caption{Architecture of the RL network with shared feature extractor, policy head, and value head.}
   \label{fig:net}
\end{figure}

The policy and value networks employ an actor-critic architecture with shared feature extraction (Figure~\ref{fig:net}), taking state $\mathbf{s}_i^t \in \mathbb{R}^{15}$ as input.

\textbf{Shared feature extractor.} The input state is processed through two hidden layers with dimensions 128 and 64. Each layer consists of a linear transformation followed by ReLU activation, layer normalization, and dropout (rate 0.1). This shared structure extracts common features useful for both policy and value estimation, reducing parameters and improving sample efficiency.

\textbf{Policy head.} The extracted features pass through an additional hidden layer (dimension 64) with ReLU, layer normalization, and dropout. The final linear layer outputs 5 logits corresponding to the action space $\mathcal{A}$. Invalid actions are masked by setting their logits to $-\infty$ before applying softmax to obtain the action probability distribution $\pi_\theta(a \mid \mathbf{s}_i^t)$, from which the agent samples actions.

\textbf{Value head.} A similar structure with a hidden layer (dimension 64) outputs a single scalar representing the estimated state value $V_\phi(\mathbf{s}_i^t)$. This value estimate is used to compute the advantage function during policy updates.

All linear layers use Xavier uniform initialization with zero biases, ensuring stable gradient flow during early training stages.

\subsubsection{Training Algorithm}

We train the agent using PPO with online updates during the attack process. The algorithm employs clipped surrogate objectives and Generalized Advantage Estimation (GAE)~\cite{gae}.

\textbf{Advantage estimation.} The temporal difference error and advantage are computed as:
\begin{align}
\delta_t &= r_t + \gamma V_\phi(\mathbf{s}_{t+1}) (1 - d_t) - V_\phi(\mathbf{s}_t) \\
A_t &= \delta_t + \gamma \lambda (1 - d_t) A_{t+1}
\end{align}
where $\gamma$ is the discount factor, $\lambda$ is the GAE parameter, and $d_t$ indicates episode termination. Returns are $G_t = A_t + V_\phi(\mathbf{s}_t)$. Advantages are normalized across batches before policy updates.

\textbf{Loss functions.} The policy loss uses clipped probability ratios $\rho_t = \pi_\theta(a_t \mid \mathbf{s}_t) / \pi_{\theta_{\text{old}}}(a_t \mid \mathbf{s}_t)$:
\begin{equation}
\mathcal{L}_{\text{policy}}(\theta) = -\frac{1}{T}\sum_{t=0}^{T-1} \min\big[\rho_t \hat{A}_t, \text{clip}(\rho_t, 1\pm\epsilon) \hat{A}_t\big]
\end{equation}
The value loss is mean squared error:
\begin{equation}
\mathcal{L}_{\text{value}}(\phi) = \frac{1}{T}\sum_{t=0}^{T-1} (G_t - V_\phi(\mathbf{s}_t))^2
\end{equation}
An entropy bonus encourages exploration:
\begin{equation}
\mathcal{L}_{\text{entropy}}(\theta) = -\frac{1}{T}\sum_{t=0}^{T-1} H(\pi_\theta(\cdot \mid \mathbf{s}_t))
\end{equation}

\textbf{Update procedure.} For each update, we perform multiple optimization epochs on collected transitions using Adam optimizer. Gradient clipping prevents explosion. After updates, the old policy is synchronized $\pi_{\theta_{\text{old}}} \leftarrow \pi_\theta$, and the experience buffer is cleared for new collection. This online learning paradigm enables the agent to continuously adapt its strategy based on the target LLM's evolving responses, discovering model-specific vulnerabilities in real-time.

\section{Evaluation}

\subsection{Experimental Setup}

\textbf{Datasets.} Following prior work, we evaluate the iMIST framework on two benchmark datasets for jailbreak attacks. JailbreakBench Behaviors (JBB)~\cite{jbb} is an open-source robustness benchmark that curates representative samples from prior work~\cite{gcg, tdc, harmbench}. The dataset contains 100 harmful behaviors categorized into 10 misuse categories defined in OpenAI's usage policies. HarmfulQA~\cite{jades} consists of 50 harmful questions with answers based on Wikipedia content. These datasets provide sufficient coverage for evaluating jailbreak effectiveness.

\textbf{Target LLMs.} We evaluate three state-of-the-art open-source LLMs with different architectures and alignment methods. DeepSeek-V3~\cite{deepseek-v3} is a mixture-of-experts model with 671B total parameters. Qwen3-32B~\cite{qwen3} is a dense model with 32B parameters. GPT-OSS-120B~\cite{gpt_oss} is a 120B parameter model. All three models employ multiple alignment techniques including SFT, RLHF, and DPO. Additionally, GPT-OSS-120B introduces deliberative alignment, which employs chain-of-thought reasoning to improve safety responses by teaching the model to refuse harmful requests through explicit reasoning steps. These models represent diverse alignment strategies and model scales, providing a comprehensive evaluation of the attack framework.

\textbf{Comparison Methods.} To ensure fair comparison, we select representative black-box jailbreak methods that do not require access to model internals. We exclude white-box methods such as GCG and AutoDAN that rely on gradient information. The evaluated methods include: \textbf{Baseline}, \textbf{ArtPrompt}, \textbf{FlipAttack}, \textbf{PAIR}, \textbf{TAP}, \textbf{DRA}, \textbf{WordGame}, \textbf{SATA}, \textbf{RL-Jack} and \textbf{PASS}. These methods have been described in detail in the related work section and represent diverse black-box attack strategies.

\textbf{iMIST Settings}: In our experiments, the iMIST framework is configured with upper limits of 5 for both the number of decomposed sub-tasks and the RL steps per sub-task. The PPO update batch size is set to 8, and the target optimization score threshold is 0.75. Following common hyperparameter settings for PPO-based RL agents, the learning rate is $0.0003$, the discount factor $\gamma$ is $0.99$, the GAE smoothing factor $\lambda$ is $0.95$, and the PPO clipping parameter $\epsilon$ is $0.2$. The value loss coefficient is $0.5$, the entropy coefficient is $0.01$, and the maximum gradient norm for clipping is $0.5$. All code is provided in supplementary materials, for more implementation details, please refer to the code.

\textbf{Evaluation Metrics.} We employ the following metrics to comprehensively evaluate the effectiveness and stealthiness of jailbreak attacks:
\begin{itemize}
  \item \textbf{JADES Score:} We employ the JADES Score~\cite{jades} as our primary evaluation metric. JADES is a standard benchmark for jailbreak evaluation that effectively assesses the harmfulness of attack outputs by decomposing a harmful question into a set of weighted sub-questions, evaluating each sub-answer, and aggregating the results into a final score.
  \item \textbf{StrongREJECT Score:} We also adopt the StrongREJECT Score~\cite{strongreject} as a complementary metric. StrongREJECT evaluates jailbreak effectiveness from three perspectives: refusal, convincingness, and specificity. This multi-dimensional assessment provides improved accuracy compared to binary classification methods.
  \item \textbf{Detection Rate:} We evaluate the stealthiness of attack methods using three LLM-based defenders. Perplexity~\cite{ppl_defense} calculates the perplexity score of input prompts, where excessively high scores indicate anomalous inputs that would be filtered. LlamaGuard~\cite{llama_guard} and ShieldLM~\cite{shieldlm} are fine-tuned models for detecting malicious content. If the LLM output is classified as unsafe by these detectors, it would be blocked in practice.
\end{itemize}

\subsection{Results and Analysis}

\begin{table*}[t]
\centering
\caption{Attack performance across different target models and datasets.}
\label{tab:main_results}
\small
\begin{tabular}{l|cc|cc|cc}
\hline
\multirow{2}{*}{Method} & \multicolumn{2}{c|}{DeepSeek-V3} & \multicolumn{2}{c|}{Qwen3-32B} & \multicolumn{2}{c}{GPT-OSS-120B} \\
\cline{2-7}
& HarmfulQA & JBB & HarmfulQA & JBB & HarmfulQA & JBB \\
\hline
Baseline & 0.06/0.04/70\% & 0.14/0.07/59\% & 0.13/0.07/42\% & 0.10/0.09/65\% & 0.00/0.00/100\% & 0.01/0.01/99\% \\
ArtPrompt & 0.24/0.31/10\% & 0.27/0.08/17\% & 0.20/0.19/16\% & 0.20/0.18/30\% & 0.06/0.17/72\% & 0.02/0.07/95\% \\
FlipAttack & 0.41/0.57/2\% & 0.26/0.10/\textbf{0\%} & 0.30/0.48/20\% & 0.16/0.34/28\% & 0.04/0.08/88\% & 0.01/0.01/99\% \\
PAIR & 0.29/0.07/4\% & 0.30/0.45/10\% & 0.35/0.22/18\% & 0.36/0.24/15\% & 0.05/0.04/58\% & 0.03/0.06/92\% \\
TAP & 0.37/0.52/8\% & 0.11/0.08/19\% & 0.36/0.41/14\% & 0.40/0.42/12\% & 0.02/0.10/84\% & 0.02/0.04/94\% \\
DRA & 0.31/0.53/18\% & 0.22/0.43/38\% & 0.28/0.25/12\% & 0.21/0.12/13\% & 0.08/0.18/78\% & 0.06/0.14/85\% \\
WordGame & 0.20/0.24/\textbf{2\%} & 0.13/0.09/49\% & 0.22/0.40/20\% & 0.24/0.40/22\% & 0.22/0.24/46\% & 0.12/0.14/65\% \\
SATA & 0.18/0.34/30\% & 0.16/0.06/44\% & 0.20/0.38/22\% & 0.15/0.22/47\% & 0.08/0.12/76\% & 0.02/0.08/90\% \\
RL-Jack & 0.17/0.28/24\% & 0.18/0.34/27\% & 0.14/0.07/40\% & 0.17/0.12/38\% & 0.02/0.04/92\% & 0.00/0.00/100\% \\
PASS & 0.24/0.27/14\% & 0.28/0.56/9\% & 0.24/0.18/8\% & 0.23/0.20/15\% & 0.10/0.16/64\% & 0.08/0.16/72\% \\
\hline
iMIST & \textbf{0.56}/\textbf{0.60}/\textbf{2\%} & \textbf{0.50}/\textbf{0.64}/\textbf{4\%} & \textbf{0.48}/\textbf{0.58}/\textbf{4\%} & \textbf{0.46}/\textbf{0.54}/\textbf{0\%} & \textbf{0.41}/\textbf{0.38}/\textbf{32\%} & \textbf{0.38}/\textbf{0.30}/\textbf{36\%} \\
\hline
\end{tabular}
\begin{tablenotes}
\item Scores are reported in the format JADES/StrongREJECT/RR, where RR denotes rejection rate.
\end{tablenotes}
\end{table*}

\textbf{Attack Effectiveness.}
Table~\ref{tab:main_results} presents attack performance across all methods. The results show that different models exhibit varying robustness against different attacks. For instance, FlipAttack achieves a JADES score of 0.41 on DeepSeek-V3 with HarmfulQA but only 0.04 on GPT-OSS-120B, while PAIR shows a JADES score of 0.29 on DeepSeek-V3 but 0.36 on Qwen3-32B. GPT-OSS-120B demonstrates stronger defense capabilities, with most baselines achieving JADES scores below 0.10, likely due to its deliberative alignment mechanism. Notably, low rejection rates do not necessarily correlate with high effectiveness, FlipAttack achieves 0\% rejection on DeepSeek-V3 with JBB but only a JADES score of 0.26, while PAIR achieves 4\% rejection with a higher score of 0.30. This occurs because some methods bypass safety mechanisms but produce responses lacking useful harmful information. iMIST consistently outperforms all baselines across both effectiveness metrics and rejection rates. On DeepSeek-V3 with HarmfulQA, iMIST achieves a JADES score of 0.56 and StrongREJECT score of 0.60, compared to the best baseline FlipAttack with 0.41 and 0.57 respectively. On Qwen3-32B with JBB, iMIST achieves 0\% rejection while maintaining a JADES score of 0.46. Even on the more robust GPT-OSS-120B, iMIST maintains a JADES score of 0.41 on HarmfulQA with only 32\% rejection rate. The interactive progressive optimization refines response harmfulness after bypassing safety filters, ensuring responses remain focused on malicious intent while providing specific and actionable information.

\begin{table}[h]
\centering
\caption{Unsafe rates by different defense methods on DeepSeek-V3 with JBB dataset.}
\label{tab:detection}
\small
\begin{tabular}{l|ccc}
\hline
Method & LlamaGuard & ShieldLM & PPL \\
\hline
Baseline & 41.46\% & 19.51\% & 86.17 \\
ArtPrompt & 91.94\% & 64.56\% & 10.03 \\
FlipAttack & 100.00\% & 96.15\% & 24.73 \\
PAIR & 44.32\% & 21.84\% & 13.75 \\
TAP & 34.67\% & \textbf{6.76\%} & 20.67 \\
DRA & 100.00\% & 92.31\% & 13.71 \\
WordGame & 70.97\% & 56.67\% & 7.24 \\
SATA & 78.57\% & 46.43\% & \textbf{5.19} \\
RL-Jack & 23.29\% & 10.96\% & 13.28 \\
PASS & 95.77\% & 81.61\% & 30.52 \\
\hline
iMIST & \textbf{17.65\%} & 14.71\% & 28.19 \\
\hline
\end{tabular}
\end{table}

\textbf{Defense Evasion.}
Table~\ref{tab:detection} presents detection rates of three defense mechanisms on DeepSeek-V3 with JBB. We evaluate only \textbf{non-rejected} responses from the main experiments. For each method, we feed input-output pairs to LlamaGuard and ShieldLM to compute unsafe detection rates, and calculate average perplexity using Llama-3.1-8B-Instruct. For iMIST, the input consists of the disguised query and target LLM output in OpenAI tool call format. iMIST maintains low detection rates of 17.65\% and 14.71\% for LlamaGuard and ShieldLM respectively, comparable to or lower than most baselines. This stealthiness is attributed to the tool disguised format, where structured tool call responses make it difficult for content-based detectors to identify harmful intent. The average PPL of 28.19 is slightly elevated due to JSON structure but remains within reasonable range.

\begin{figure}[h]
   \centering
   \includegraphics[width=1.0\linewidth]{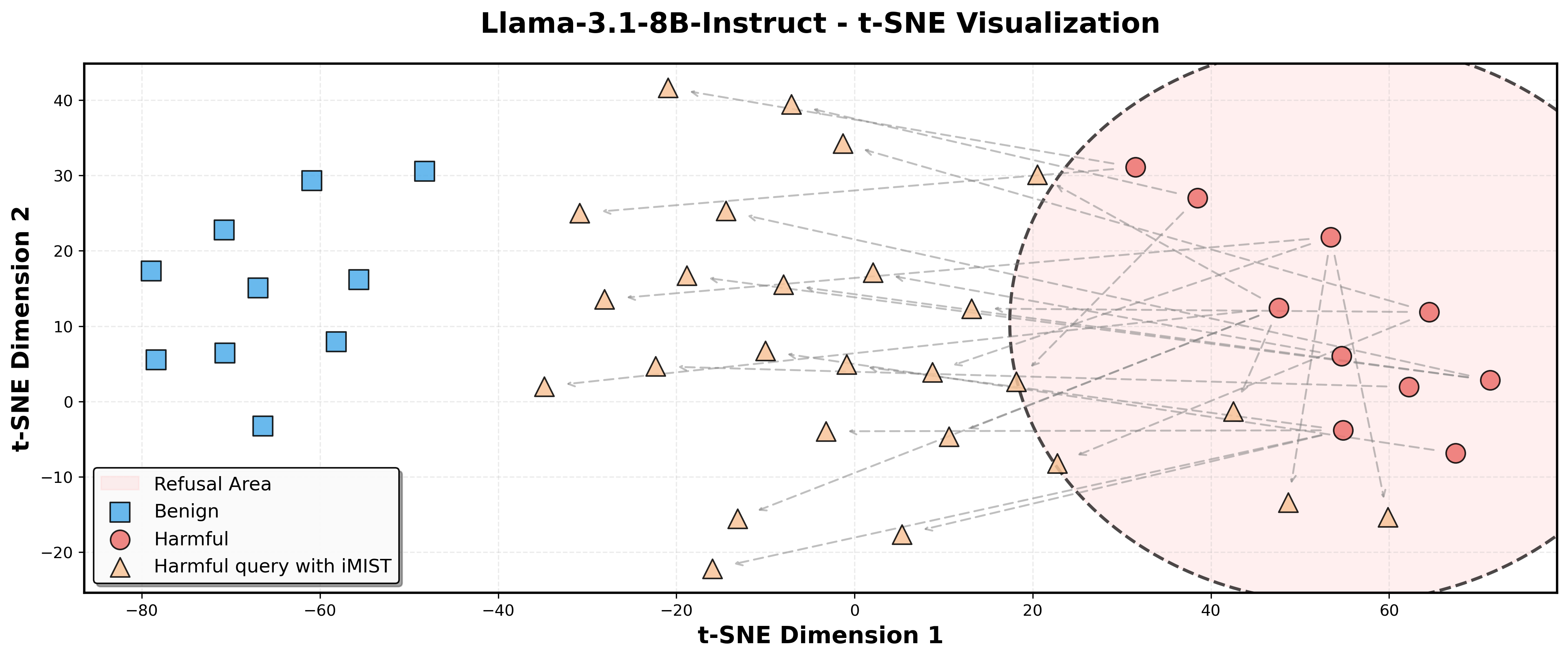}
   \caption{t-SNE visualization of hidden representations in Llama-3.1-8B-Instruct.}
   \label{fig:tsne}
\end{figure}

\textbf{Representation Analysis.}
To understand how iMIST works, we follow prior work~\cite{boost, pass} to visualize the hidden representation space of aligned models. We randomly sample 10 harmful queries from HarmfulQA and generate corresponding benign prompts with minimal word changes. We then apply iMIST to attack these queries and collect the inputs sent to the target LLM. For each prompt type (benign, harmful, and iMIST-attacked), we extract the last token's hidden representation from the -10th layer of Llama-3.1-8B-Instruct, and visualize them using t-SNE 2D projection.
Figure~\ref{fig:tsne} reveals a clear separation in the representation space. Benign prompts cluster in the left region, while harmful prompts concentrate in the right region within the refusal area (dashed circle). Notably, iMIST-attacked queries escape from the refusal area and are distributed across the middle region, positioning between benign and harmful clusters. The arrows illustrate how iMIST transforms harmful queries to move them away from the refusal area. This intermediate positioning suggests iMIST retains malicious intent while reducing similarity to patterns in the alignment training corpus, thereby lowering detection probability by safety mechanisms. However, some iMIST samples still fall within or near the refusal area, indicating they may be recognized by alignment defenses. This explains why interactive progressive optimization is necessary to further refine the attack strategy through iterative adjustments.

\section{Conclusion}

In this paper, we propose iMIST, a novel jailbreak attack method that exploits the tool-use capabilities of large language models through Tool Disguised Invocation and Interactive Progressive Optimization. Our evaluation demonstrates the effectiveness of iMIST across multiple target models and datasets. The results also showcase the potential of reinforcement learning in guiding LLM conversations and refining attack strategies. These findings highlight the vulnerability in current LLM safety and underscore the need for more robust defenses that account for these attacks.

\bibliographystyle{IEEEtran}
\bibliography{main}

\end{document}